\documentclass[conference]{IEEEtran}
\usepackage[cmex10]{amsmath}
\usepackage{amsfonts}
\usepackage{graphicx}
\usepackage[all]{xy}
\usepackage{booktabs}
\usepackage{verbatim}

\newtheorem{theorem}{Theorem}[section]
\newtheorem{definition}[theorem]{Definition}

\newtheorem{corollary}[theorem]{Corollary}
\newtheorem{proposition}[theorem]{Proposition}
\newtheorem{lemma}[theorem]{Lemma}

\newtheorem{remark}[theorem]{Remark}
\newcommand{\ff}{{\mathbb F}}
\newcommand{\CC}{{\mathbb C}}
\newcommand{\ZZ}{{\mathbb Z}}
\newcommand{\NN}{{\mathbb N}}
\newcommand{\F}{\mathbb F}

\def\Tr{\mathop{{\rm Tr}}}

\def\CK{\mathcal{K}}

\def\Res{\mathop{\operatorfont Res}\nolimits}

\newcommand{\cE}{\mathcal E}

\def\wt{\mathop{{\rm wt}}}

\def\uu{{\mathbf{u}}}

\def\v{{\mathbf{v}}}
\def\cc{{\mathbf{c}}}
\def\d{{\mathbf{d}}}

\def\aa{{\mathbf{a}}}
\def\b{{\mathbf{b}}}
\def\0{{\mathbf{0}}}
\def\1{{\mathbf{1}}}

\def\CduTr8E{{C^{\perp_{\Tr_{8/2} \E}}}}

\def\Tr{\mathop{{\rm Tr}}}
\def\H{\mathop{{\rm H}}}
\def\E{\mathop{{\rm E}}}

\def\bvar{{\boldsymbol{\varphi}}}
\def\bpsi{{\boldsymbol{\psi}}}

\usepackage{url}
\urldef{\radoemail}\path|{rkirov}@ntu.edu.sg|
\urldef{\fredemail}\path|{frederic.ezerman}@gmail.com|

\begin{document}
\title{Nonbinary Quantum Codes from Two-Point Divisors on Hermitian Curves}
\author{Martianus Frederic Ezerman and Radoslav Kirov%
\thanks{M.~F.~Ezerman and R.~Kirov are with the School of Physical and Mathematical Sciences, Nanyang Technological University, Singapore 631317, Republic of Singapore email: \kirovemail.}}%

\maketitle
\begin{abstract}
Sarvepalli and Klappenecker showed how classical one-point codes on the Hermitian curve can be used to construct quantum codes.
Homma and Kim determined the parameters of a larger family of codes, the two-point codes. In quantum error-correction, 
the observed presence of asymmetry in some quantum channels led to the study of asymmetric quantum codes (AQECCs) where we 
no longer assume that the different types of errors are equiprobable.
This paper considers quantum codes constructed from the two-point codes. In the asymmetric case, we show strict improvements 
over all possible finite fields for a range of designed distances. We produce large dimension pure AQECC and small dimension impure AQECC that have better parameters than AQECC from one-point codes. 
Numerical results for the Hermitian curves over $\ff_{16}$ and $\ff_{64}$ are used to illustrate the gain. 
\end{abstract}

\section{Introduction}
The term \textit{quantum codes} is a shorthand for quantum error-correcting codes. 
Quantum codes have been garnering a lot of interest since they protect 
information-carrying quantum states against decoherence and play an 
important part in making fault-tolerant quantum computation possible.
Quantum codes can be distinguished into \textit{pure} and 
\textit{impure} (or \textit{degenerate}). The pure ones are usually easier 
to implement due to their simpler decoding process while the degenerate ones 
give us better error-detecting capabilities.

In the nonbinary cases, a firm connection between classical 
error-correcting codes and quantum codes is well-established. 
We can construct quantum codes from classical codes by using 
the stabilizer formalism~\cite{KKKS06}. The resulting 
quantum codes are called \textit{stabilizer codes}. A subclass of 
these codes can be derived by using the CSS method.

The class of Hermitian codes is known to have excellent parameters. 
They are easy to describe, to encode and to decode. The most studied 
Hermitian codes are the one-point codes. Vector spaces of functions 
that correspond to two-point divisors were first studied in~\cite{Mat01}. 
A complete description of the minimum distances of all two-point 
Hermitian codes is given in~\cite{HomKim06}. Further results discussed in~\cite{HomKim06},
\cite{Bee07},~\cite{Park10}, and~\cite{DuuKir10} improve our understanding of these codes. 
Two-point codes have better parameters than one-point codes, while maintaining 
their ease of construction.

This paper is organized as follows. Section~\ref{sec:prelim} contains three subsections. 
They discuss, respectively, Hermitian codes, quantum codes, and three relevant 
construction methods that will be needed to derive quantum codes. 
Section~\ref{sec:QcodesHcurve} establishes the parameters of quantum codes derived from 
two-point Hermitian codes and compare them with the corresponding parameters of 
quantum codes from one-point codes. Using coset bounds we can construct excellent impure 
AQECCs of small dimension. This fact and related results are contained in Section~\ref{sec:asym}.

\section{Preliminaries}\label{sec:prelim}
Let $\F_{q}$ denote the finite field of cardinality $q = p^{m}$ 
for a prime $p$ and $m \in \NN$. The trace mapping 
$\Tr: \F_{q} \to \F_{p}$ is given by $\Tr (\beta) = 
\beta + \beta^{p}+\beta^{p^{2}}+\ldots+\beta^{p^{m-1}}$.
Given any two distinct (nonempty) subsets $C$ and $D$ of $\F_{q}^{n}$, 
let the notation $\wt(C \setminus D)$ denote $\min \{ \wt(\uu) :
\uu\in(C \setminus D),~\uu \neq \0\}$ with $\wt(\uu)$ denoting the 
Hamming weight of $\uu$.

For $\uu=(u_1,u_2,\ldots,u_n),\v=(v_1,v_2,\ldots,v_n) \in \F_{q}^n$,
\begin{enumerate}
 \item $\langle \uu, \v \rangle_{\E}=\sum_{i=1}^{n} u_i v_i$ is
the \textit{Euclidean inner product} of $\uu$ and $\v$.
 \item If $\F_{q}$ is a quadratic extension of $\F_{e=p^{l}}$, then
$\langle \uu, \v \rangle_{\H}= \sum_{i=1}^{n} u_i v_i^{e}$ is
the \textit{Hermitian inner product} of $\uu$ and $\v$.
\end{enumerate}

Let $C$ be an $[n,k,d]_{q}$-code. Let $*$ represent either the Euclidean or 
the Hermitian inner product, the \textit{dual code}
$C^{\perp_{*}}$ of $C$ is given by
\begin{equation*}
 C^{\perp_{*}} := \left\lbrace \uu \in \F_{q}^{n} : \left\langle
\uu,\v\right\rangle _{*} = 0 \text{ for all } \v \in C
\right\rbrace
\end{equation*}
while the dual distance $d^{\perp_{*}}$ is defined to be $d(C^{\perp_{*}})$.

A \textit{monomial matrix} $M$ is a square matrix over $\F_{q}$ with exactly
one nonzero entry in each row and each column. Such a matrix can be written 
as $TP$ or $PT'$ where $T$ and $T'$ are diagonal matrices and $P$ is a
permutation matrix. Two codes $C$ and $C'$ are said to be
\textit{(monomially) equivalent} if there is a monomial matrix $M$ such that
$G'=M G$, for the corresponding generator matrices $G$ and $G'$.  Equivalent
codes have the same parameters.

\subsection{Hermitian Codes}\label{subsec:Hcodes}
We recall Goppa's general construction of codes from curves. Let $X/\ff_{q}$ be an 
algebraic curve (absolutely irreducible, smooth, projective) of genus $g$ over $\ff_{q}$. 
Let $\ff_{q}(X)$ be the function field of $X/\ff_{q}$ and $\Omega(X)$ be the module of 
rational differentials of $X/\ff_{q}$.

Given a divisor $E$ on $X$ defined over $\ff_{q}$, let $L(E) = 
\{ f \in \ff_{q}(X) \setminus \{0\} : (f)+E\geq 0 \} \cup \{0\},$ and 
let $\Omega(E) = \{ \omega \in \Omega(X) \setminus\{0\} : (\omega) \geq E \} \cup \{0\}.$ 
Let $K$ represent the canonical divisor class. For $n$ distinct rational points 
$P_1, \ldots, P_n$ on $X/\ff_{q}$ and for disjoint divisors $D=P_1+\cdots+P_n$ and $G$, 
the geometric Goppa codes $C_L(D,G)$ and $C_\Omega(D,G)$ are defined as the images of the maps 
\begin{align*}
\alpha _L :~ &L(G)~\longrightarrow~\ff^{\,n}, ~f \mapsto (f(P_1), \ldots, f(P_n) ). \\
\alpha _\Omega :~ &\Omega(G-D)~\longrightarrow~\ff^{\,n}, \\
                  &\omega \mapsto (\Res_{P_1}(\omega), \ldots, \Res_{P_n}(\omega)).
\end{align*}

A consequence of the Residue Theorem for function fields is that $C_L(D,G)^{\perp_{\E}} 
= C_\Omega(D,G)$ \cite[Theorem 2.2.8]{Sti09}. 
Moreover, the residue construction can be represented as an evaluation. 
\begin{lemma}[{\cite[Proposition 2.2.10]{Sti09}}] \label{L:evres}
  Let $\nu$ be a differential with simple poles and residue $1$ at the points of $D$. Then
  \[
C_L(D, G)^{\perp_{\E}} = C_\Omega(D, G) = C_L(D, D - G + (\nu)).
\]
\end{lemma}

In this paper we only consider the Hermitian curve, 
which is the smooth projective curve over $\ff_{q^2}$ with affine equation 
$y^q + y = x^{q+1}$. It achieves the Hasse-Weil bound with $q^3 + 1$ rational points 
and genus $g = q(q-1)/2$. 

Classical two-point codes are the codes $C_L(D,G)$ and $C_\Omega(D,G)$ with Goppa divisor $G=iP+jQ$. 
To construct them, we fix two distinct rational points $P$ and $Q$. 
The standard choice is to let $P$ be the point at infinity (the common pole of $x$ and $y$)
and $Q$ be the origin (the common zero of $x$ and $y$). The equivalent divisors $(q+1)P \sim (q+1)Q$ 
belong to the hyperplane divisor class $H$ with respect to the model above.
The divisor sum $R$ of all $q^3+1$ rational points belongs to the divisor class $(q^2-q+1)H$ 
and the canonical divisor class $K=(q-2)H$. See~\cite{Tie87},~\cite{Sti88},~\cite[Section 8.3]{Sti09} and~\cite[Section 4.4.3]{TsfVla07} for the details.

Henceforth, we fix the divisor $D$ to be $R - P - Q$, making the length of the constructed codes $q^3 -1$. 
The two-point codes are one coordinate shorter than the one-point codes.
In order to compare the two families, we shorten one-point codes.
Since the automorphism group of a one-point code acts transitively on the set of coordinates~\cite[Remark 8.3.6]{Sti09}, 
the choice of coordinate is non-essential. 
Thus the minimum distance of the code is preserved under shortening. This feature makes it 
easy to compare two-point codes of length $q^3-1$ with the shortened one-point codes of equal dimension.

It is known that the Euclidean duals of one-point codes are also one-point codes. We extend this property to two-point codes.
\begin{proposition} \label{P:dual}
  If $D = R - P - Q$, then
\begin{multline*}
C_L(D, iP - jQ)^{\perp_{\E}} = \\ 
C_L(D, (q^3 + q^2 - q - 2 - i)P + (j-1)Q). 
\end{multline*}
\end{proposition}
\begin{IEEEproof}
 Following the proof for one-point codes in~\cite[Proposition 8.3.2]{Sti09}, 
we select $\nu = dt/t$, with $t = x^{q^2} - x$, and apply Lemma \ref{L:evres}.
\end{IEEEproof}

Self-orthogonality property is important in some construction of quantum codes.
\begin{corollary} \label{C:cont}
If $D = R - P - Q$, $G = iP + jQ$ and $G' = i'P + j'Q$, then 
  \[
C_L(D, G)^{\perp_{\E}} \subseteq C_L(D, G') 
\]
if $q^3 + q^2 - q - 2 \leq i + i'$ and $-1 \leq j + j'$. 
The code $C_L(D, iP + jQ)$ is Euclidean self-orthogonal if $ 2i \leq q^3 + q^2 - q - 2$ and $j < 0$.
\end{corollary}

Equivalent divisors $G \sim \widehat{G}$ produce equivalent geometric Goppa codes 
$C(D,G)$ and $C(D,\widehat{G})$ under both maps $\alpha_L$ and $\alpha_\Omega$.

Due to the equivalence $(q+1)P \sim (q+1)Q$, every two-point code (using either construction) 
is uniquely equivalent to a code of the form $C_L(D, iP - jQ)$ with $0 \leq j \leq q$. 
We use this representation as a canonical one. 
Moreover, a particularly favorable feature of the Hermitian curves is that one can 
explicitly write a monomial basis for the Riemann-Roch space of a two-point divisor 
of that form. 
\begin{lemma}[\cite{Park10}]\label{L:exp2pt}
Let $D = c(q+1)P - aP - bQ,$ for $c \in \ZZ,$ and for $0 \leq a,b \leq q$. 
The space $L(D)$ has a basis given by the monomials $x^{i}y^{j}$ where:
\begin{enumerate}
\item $0 \leq i \leq q,$ $0 \leq j,$ and $i+j \leq c$,
\item $a \leq i \text{ for } i+j = c$,
\item $b \leq i \text{ for } j = 0$.
\end{enumerate}
\end{lemma}
 
The actual minimum distance of two-point codes was determined by Homma and Kim in~\cite[Th. 5.2 and Th. 6.1]{HomKim05} 
for $n=0$ and $n=q$, as well as in~\cite[Th. 1.3 and Th. 1.4]{HomKim06} for $0 < n < q$. Using order bound techniques, 
Beelen in~\cite[Th. 17]{Bee07} gives lower bounds for the cases $\deg G > \deg K$ (i.e. for $m+n > (q-2)(q+1)$), 
and Park settles all cases in~\cite[Th. 3.3 and Th. 3.5]{Park10}. Park moreover shows that the lower bounds are 
sharp and that they correspond to the actual minimum distance. In~\cite{DuuKir10}, Duursma and Kirov show that among 
all divisors $G=iP+jQ$ of a given degree, the optimal minimum distance is attained for a choice of the form $G=aP-2Q$.

\begin{proposition}[\cite{DuuKir10}] \label{P:dist}
Let $G=K+B$ where $B$ is a divisor such that $B \neq 0$, $\deg B \geq 0$ and 
$B = cH -aP -qQ,$ for $0 \leq a \leq q$. If $D \cap \{P,Q\} = \emptyset$, then 
the two-point code $C_L(D,G)$ has dimension $\deg G - g + 1$ 
and dual distance
\begin{equation*}
\d^{\perp_{\E}}=
\begin{cases}
	\deg B + \max(0, q - c) \text{ if } a = q, \text{ otherwise,}\\
	\deg B + \max(0, q - c) + \max(0, a - c) \text{.}
\end{cases}
\end{equation*}
The corresponding one-point code of the same dimension has the same minimum distance if $a = q$, but $\max(0,q-c)$ less if otherwise.
\end{proposition}

Using Proposition \ref{P:dual}, we can restate the result for the minimum distance of the evaluation codes.
\begin{corollary} \label{C:best}
Let $0 \leq r \leq q(q+1)$, and let $1 \leq c$ and $0 \leq a \leq q$ be the unique numbers such that $r + q = c(q+1) - a$.
The code $C_L(D, (q^3 - r + 1)P  - 2Q)$ is a $[q^3-1,k(r),d(r)]_{q^2}$ code where 
\begin{align*}
k(r) &= q^3 - q(q-1)/2 - r, \\
d(r) &=
\begin{cases}
	r + \max(0, q - c) \text{ if } a = q, \text{ otherwise,}\\
	r + \max(0, q - c) + \max(0, a - c) \text{.}
\end{cases}
\end{align*}
\end{corollary}

Note that the range for $r$ can be extended, but outside the given range there are no improvements over one-point codes.
\subsection{Quantum Codes}\label{subsec:Qcodes}
Let $\CC$ be the field of complex numbers and
$\eta=e^{\frac{2\pi \sqrt{-1}}{p}}\in \CC$.
Let $V_{n}=(\CC^{q})^{\otimes n }=\CC^{q^{n}}$
be the $n$th tensor product of $\CC^{q}$.
$V_{n}$ has the following orthonormal basis
\begin{equation}\label{basis}
\left\{|\cc\rangle =|c_{1} c_{2}\ldots c_{n}\rangle
: \cc=(c_{1},\ldots,c_{n}) \in \F_{q}^n\right\} \text{,}
\end{equation}
where $|c_{1} c_{2}\ldots c_{n}\rangle$ abbreviates
$|c_{1}\rangle\otimes|c_{2}\rangle\otimes\cdots \otimes
|c_{n}\rangle$.

For two quantum states $|\bvar\rangle$ and $|\bpsi\rangle$ in 
$V_{n}$ with 
\begin{equation*}
|\bvar\rangle=\sum\limits_{\cc\in \F_{q}^{n}}\alpha(\cc)|\cc\rangle 
\text { and } |\bpsi\rangle=\sum\limits_{\cc\in \F_{q}^{n}}\beta(\cc)|\cc\rangle \text{,}
\end{equation*}
where $\alpha(\cc),\beta(\cc)\in \CC$, the inner product of 
$|\bvar\rangle$ and $|\bpsi\rangle$ is given by
\begin{equation*}
\langle \bvar|\bpsi\rangle=\sum\limits_{\cc\in \F_{q}^{n}}
\widetilde{\alpha(\cc)}\beta(\cc)\in \CC \text{,} 
\end{equation*}
where  $\widetilde{\alpha(\cc)}$ is the complex conjugate of 
$\alpha(\cc)$. We say $|\bvar\rangle$ and $|\bpsi\rangle$ are 
\textit{orthogonal} if $\langle \bvar|\bpsi\rangle=0$.

Essentials on the standard mathematical error model
of quantum error-correction can be found, for instance, in~\cite{AK01} and
in~\cite{KKKS06} for the symmetric case and in~\cite{WFLX09} for the
asymmetric case.

To define a quantum code $Q$, we need to consider the
set of error operators that $Q$ can handle. Let $\alpha, \beta \in \F_{q}$. The
unitary operators $X(\alpha)$ and $Z(\beta)$ on $\CC^{q}$ are defined by
\begin{equation}\label{eq:ErOp}
X(\alpha) |\varphi \rangle = |\varphi+\alpha \rangle \text{ and } Z(\beta)|\varphi\rangle =
\eta^{\Tr \left( \langle \beta,\varphi \rangle_{\E}\right)} |\varphi\rangle \text{.}
\end{equation}
Based on Equation (\ref{eq:ErOp}), for $\aa=(\alpha_1,\ldots,\alpha_{n}) \in \F_{q}^{n}$,
we can write $X(\aa) = X(\alpha_{1}) \otimes \ldots \otimes X(\alpha_{n})$ and
$Z(\aa) = Z(\alpha_{1}) \otimes \ldots \otimes Z(\alpha_{n})$ for the tensor product of
$n$ error operators. The set $\cE_{n}:=\{ X(\aa)Z(\b) : \aa,\b \in \F_{q}^{n} \}$
is a nice error basis on $V_{n}$.

The error group $G_{n}$ of order $pq^{2n}$ is generated by the matrices in $\cE_{n}$
\begin{equation*}
 G_{n}:=\{ \eta^{c} X(\aa) Z(\b) : \aa,\b \in \F_{q}^{n}, c \in \F_{p} \} \text{.}
\end{equation*}
Let $E = \eta^{c} X(\aa) Z(\b) \in G_{n}$. Then the \textit{quantum weight}
${\rm wt_{Q}}(E)$ of $E$ is the number of coordinates such that 
$(\alpha_{i},\beta_{i}) \neq (0,0)$. The number of $X$-errors ${\rm wt_{X}}(E)$
and the number of $Z$-errors ${\rm wt_{Z}}(E)$ in the error operator $E$ are given,
respectively, by $\wt (\aa)$ and $\wt (\b)$.

\begin{definition}\label{def:AQECC}
A \textit{$q$-ary  quantum code} of length $n$ is a subspace $Q$ of
$V_{n}$ with dimension $\CK \geq1$. A quantum code $Q$ of dimension
$\CK\geq2$ is said to detect $d-1$ quantum digits of errors for
$d\geq1$ if, for every orthogonal pair $|\bvar\rangle$ and $|\bpsi\rangle$ in
$Q$ and every $E \in G_{n}$ with ${\rm wt_{Q}}(E) \leq d-1$, 
$|\bvar\rangle$ and $E|\bpsi\rangle$ are orthogonal. 
In this case, we call $Q$  a \textit{symmetric}
quantum code with parameters $((n,\CK,d))_{q}$ or
$[[n,k,d]]_{q}$, where $k=\log_{q} \CK$. Such a quantum code is called
\textit{pure} if $|\bvar\rangle$ and $E|\bpsi\rangle$ are orthogonal for 
any (not necessarily orthogonal) $|\bvar\rangle$ and
$|\bpsi\rangle$ in $Q$ and any $E \in G_{n}$ with $1 \leq {\rm wt_{Q}}(E) \leq
d-1$. A quantum code $Q$ with $\CK=1$ is assumed to be pure.

Let $d_{x}$ and $d_{z}$ be positive integers. A
quantum code $Q$ in $V_{n}$ with dimension $\CK\geq2$ is called an 
\textit{asymmetric quantum code} with parameters $((n,\CK,d_{z}/d_{x}))_{q}$
or $[[n,k,d_{z}/d_{x}]]_{q}$, where $k=\log_{q} \CK$, if $Q$ detects $d_{x}-1$
quantum digits of $X$-errors and, at the same time, $d_{z}-1$
quantum digits of $Z$-errors. That is, if $\langle \bvar|\bpsi \rangle=0$ for
$|\bvar\rangle,|\bpsi\rangle\in Q$, then $|\bvar\rangle$ and $E|\bpsi\rangle$ 
are orthogonal for any $E \in G_{n}$ such that ${\rm wt_{X}}(E)\leq d_{x}-1$ 
and ${\rm wt_{Z}}(E)\leq d_{z}-1$. 
Such an asymmetric quantum code $Q$ is called \textit{pure} if $|\bvar\rangle$ 
and $E|\bpsi\rangle$ are orthogonal for any $|\bvar\rangle,|\bpsi\rangle\in Q$ and
any $E \in G_{n}$ such that $1 \leq {\rm wt_{X}}(E)\leq d_{x}-1$ or 
$1 \leq {\rm wt_{Z}}(E)\leq d_{z}-1$. An asymmetric quantum code $Q$ with $\CK=1$ 
is assumed to be pure.
\end{definition}

\begin{remark}\label{rem2.3}
An asymmetric quantum code with parameters $((n,\CK,d/d))_{q}$
is a symmetric quantum code with parameters $((n,\CK,d))_{q}$,
but the converse is not true since, for $E\in G_{n}$ with
${\rm wt_{X}}(E)\leq d-1$ and ${\rm wt_{Z}}(E)\leq d-1$, ${\rm wt_{Q}}(E)$ may be
bigger than $d-1$.
\end{remark}

\subsection{Constructions of Quantum Codes from Classical Codes}\label{subsec:construct}
It is well-known that quantum codes can be constructed from classical codes. 
We will use the following three constructions tailored to Hermitian codes.

\begin{lemma}(CSS Construction)\label{lem:CSSEuclidean}\cite[Lem. 20]{KKKS06}
	Let $C_{i}$ be an $[n,k_{i},d_{i}]_{q^{2}}$-code for $i=1,2$. Let $C_{1}^{\perp_{\E}} 
	\subseteq C_{2}$. Then there exists a symmetric quantum code $Q$ with parameters 
	$[[n,k_{1}+k_{2}-n,\min \{\wt(C_{2} \setminus C_{1}^{\perp_{\E}}),\wt(C_{1} 
	\setminus C_{2}^{\perp_{\E}}) \} ]]_{q^{2}}$ which is pure whenever 
	$\min \{\wt(C_{2} \setminus C_{1}^{\perp_{\E}}),\wt(C_{1} 
	\setminus C_{2}^{\perp_{\E}}) \} = \min \{ d_{i}\}$.
	If we have $C \subseteq C^{\perp_{\E}}$ where $C$ is an $[n,k,d]_{q^{2}}$-code, then $Q$ 
	is an $[[n,n-2k,\wt(C^{\perp_{\E}} \setminus C)]]_{q^{2}}$-code which is pure 
	whenever $d^{\perp_{\E}}= \wt(C^{\perp_{\E}} \setminus C)$.
\end{lemma}

If, instead of the Euclidean, we use the Hermitian inner product, we have 
the following construction of a $q$-ary quantum code from a Hermitian self-orthogonal 
code $C \subseteq \F_{q^{2}}^{n}$.
\begin{lemma}\cite[Cor. 19]{KKKS06}
  \label{L:herm}
	Let $C$ be an $[n,k,d]_{q^{2}}$-code such that $C \subseteq C^{\perp_{\H}}$. 
	Then there exists a symmetric quantum code $Q$ with parameters 
	$[[n,n-2k,\wt(C^{\perp_{\H}} \setminus C)]]_{q}$-code which is pure whenever 
	$\wt(C^{\perp_{\H}} \setminus C) = d^{\perp_{\H}}$.
\end{lemma}

The CSS construction extends to the AQECCs derived from Hermitian codes. 
We can use either the Euclidean or the Hermitian inner product if $q=e^{2}$ 
since, over $\F_{q^{2}}$, $C^{\perp_{\E}}$ and $C^{\perp_{\H}}$ share the 
same MacWilliams transform, making $d^{\perp_{\E}} = d^{\perp_{\H}}$.

\begin{lemma}\cite[Lem. 3.1]{SKR09}
  \label{L:cssass}
	Let $C_{i}$ be an $[n,k_{i},d_{i}]_{q^{2}}$-code for $i=1,2$. Let $C_{1}^{\perp_{*}} 
	\subseteq C_{2}$. Let $d_{z}:=\max \{ \wt(C_{2} \setminus C_{1}^{\perp_{*}}), \wt (C_{1} 
	\setminus C_{2}^{\perp_{*}}) \}$ and $d_{x}:=\min \{ \wt(C_{2} \setminus C_{1}^{\perp_{*}}),
	\wt(C_{1} \setminus C_{2}^{\perp_{*}}) \}$. Then there exists an asymmetric quantum code 
	$Q$ with parameters $[[n,k_{1}+k_{2}-n,d_{z}/d_{x}]]_{q^{2}}$. The code $Q$ is pure whenever 
	$\{ d_{z},d_{x} \} = \{ d_{1},d_{2} \}$.
	If we have $C \subseteq C^{\perp_{*}}$ where $C$ is an $[n,k,d]_{q^{2}}$-code, then $Q$ is an 
	$[[n,n-2k,d'/d']]_{q^{2}}$-code where $d' = \wt(C^{\perp_{*}} \setminus C)$. 
	The code $Q$ is pure whenever $d'=d^{\perp_{*}}$.
\end{lemma}

\section{Quantum Codes from Hermitian Curve}\label{sec:QcodesHcurve}
We apply Lemmas~\ref{L:herm} and~\ref{L:cssass} to construct quantum 
codes. We restrict our attention to the range where two-point codes improve on one-point codes as given in Proposition \ref{P:dist}.

First we use the CSS construction with $C_1$ and $C_2^{\perp_{\E}}$ in the range of 
improvement. This construction produces long quantum codes with excellent parameters.

\begin{proposition}\label{P:bigcss}
Let $0 \leq r_1 \leq r_2 \leq q(q+1)$. For $q \geq 4$, there exists a pure AQECC
with parameters $[ [ q^3 - 1, q^3 - q(q-1) - (r_1 + r_2) + 1, d(r_2) / d(r_1) ] ]_{q^{2}}$ where $d(r_{1})$ and $d(r_{2})$ 
are computed according to Corollary~\ref{C:best}. 
\end{proposition}
\begin{IEEEproof}
  Apply Lemma~\ref{L:cssass} with $C_1 = C_L(D, (q^3-r_1 + 1 - (q+1))P + (q-1)Q)$ and $C_2 = C_L(D, (q^3-r_2 + 1 )P - 2Q)$. 
  The nestedness $C_1^{\perp_{\E}} \subseteq C_2$ is guaranteed by Corollary~\ref{C:cont}, given the range $r_i \leq q(q+1)$.
  The minimum distance can be computed from Corollary \ref{C:best} since $C_1$ is equivalent to a code having a divisor 
  of the form $G = iP - 2Q$. 
  By the Riemann-Roch Theorem, $d \left(C_i^{\perp_{\E}}\right) = q^3 - 1 - r - (q-2)(q+1)$. If $q \geq 4$, this value is 
  larger than $d(r_i)$ for the given range. Thus the derived quantum code is pure. 
\end{IEEEproof}

Let the designed distance $\delta$ be fixed. Tables~\ref{Tab:classicalf16} and~\ref{Tab:classicalf64} 
list down the best dimension obtainable from one-point and two-point codes based on Proposition~\ref{P:dist}, 
along with the design parameter $r$ used in Corollary \ref{C:best}. 
 
By Proposition \ref{P:bigcss}, the inner and the outer codes can be independently 
selected to be optimal when constructing an AQECC, as long as they 
are within the specified range. This effectively doubles the gain when switching 
to two-point codes. For example, the best $16$-ary AQECC with $d_z = 9$ and 
$d_x = 5$ we can construct is of parameters $[[ 63, 39, 9 / 5]]_{16}$ if only 
one-point codes are considered. Using two-point codes, the value of $k$ increases 
to $42$.

\begin{table}
  \centering
  \begin{tabular}{@{}cccccc@{}}
    \toprule
    $\delta$ & \multicolumn{2}{c}{Dimension} & r\\
           &1-point & 2-point & \\
\midrule
5   &   53  &   55 & 3 \\
7   &   52  &   53 & 5 \\
9   &   49  &   50 & 8 \\
11  &   47  &   48 & 10 \\
\bottomrule
  \end{tabular}
  \caption{Dimensions of one- and two-point codes on the Hermitian curve over $\ff_{16}$} 
\label{Tab:classicalf16}
\end{table}

\begin{table}
  \centering
  \begin{tabular}{@{}cccc||cccc@{}}
    \toprule
    $\delta$ & \multicolumn{2}{c}{Dimension}&r & $\delta$ & \multicolumn{2}{c}{Dimension}& r\\
             & 1-point & 2-point          &   &         & 1-point & 2-point& \\
\midrule
9 &  475&  481&  3 & 33&  451&  454& 30\\
11&  474&  481&  3 & 35&  449&  454& 30\\
13&  474&  481&  3 & 37&  447&  450& 34\\
15&  474&  475&  9 & 39&  447&  448& 36\\
17&  467&  472&  12& 41&  443&  445& 39\\
19&  465&  472&  12& 43&  441&  443& 41\\
21&  465&  472&  12& 45&  439&  441& 43\\
23&  465&  466&  18& 47&  438&  439& 45\\
25&  459&  463&  21& 49&  435&  436& 48\\
27&  457&  463&  21& 51&  433&  434& 50\\
29&  456&  459&  25& 53&  431&  432& 52\\
31&  456&  457&  27& 55&  429&  430& 54\\

\bottomrule
  \end{tabular}
  \caption{Dimensions of one- and two-point codes on the Hermitian curve over $\ff_{64}$} 
\label{Tab:classicalf64}
\end{table}

Lemma~\ref{L:herm} states a different construction, which gives AQEEC over $\ff_{q}$ instead of $\ff_{q^2}$.
To use the construction we need the following result about the dual codes with respect to the Hermitian 
inner product. The one-point version of the proposition was proved in~\cite{SarKla06}.
\begin{proposition}
A two-point code $C_L(D,iP - jQ)$ with $1 \leq j $ is Hermitian self-orthogonal if $i \leq q^2 - 2$. 
\end{proposition}
\begin{IEEEproof}
  It is enough to prove the theorem with $j=1$ since $C_L(D,iP - jQ) \subseteq C_L(D,iP - Q)$.
By Lemma~\ref{L:exp2pt} we know that a basis for the two-point vector space can
be obtained by monomial evaluation. Codewords which are Hermitian dual to
$x^ay^b(P)$ are Euclidian dual to words of the form $x^{qa}y^{qb}(P)$ which
live in $C_L(D,qiP)$. Adding $-Q$ to the divisor removes only the constants 
and any non-constant monomial to the $q$-th power is also non-constant. Thus
the Hermitian dual of $C_L(D,iP - Q)$ contains $C_L(D,qiP - Q)^{\perp_{\E}}$.
Under the degree assumption we can use Corollary~\ref{C:cont} to show that
$C_L(D,qG)$ is Euclidean self-orthogonal.  Hence the original code $C_L(D,G)$
is Hermitian self-orthognal.  \end{IEEEproof}

Unfortunately, this requirement is too restrictive on the range of $G$.  Due to
the small degree of $G$, the dual code is outside the range of improvements
given in Proposition~\ref{P:dist}.  Thus, for this particular construction,
two-point codes do not improve on one-point codes already treated
in~\cite{SarKla06}. 
 
\section{Impure AQECCs and Coset Bounds}\label{sec:asym} Recent results
concerning the bounds for the minimum distance produce better bounds for the
cosets~\cite{DuuKirPar10}. A particular feature of the coset bounds on the
Hermitian curve is that they are non-monotonic.  This lack of monotonicity can
be exploited to produce excellent impure AQECCs based on the CSS construction. 

For the Hermitian curve over $\ff_{64}$ we simulated all possible 
pairs of two-point divisors (up to equivalence) $G_1 \leq G_2$ of degrees $0
\leq \deg G_1 \leq \deg G_2 \leq q(q-1)$ and calculated the parameters of the
impure asymmetric quantum code constructed based on the nested pair $C_1 =
C_\Omega(D,G_2) \subseteq C_\Omega(D,G_1) = C_2$. 

An implementation of the methods for coset bounds given in~\cite{DuuKirPar10}
was used to calculate $d_z = \wt(C_2 \setminus C_1)$. Note that $d_x =
\wt(C_1^{\perp_{\E}} \setminus C_2^{\perp_{\E}}) = n - \deg G_1 + 2g + 2$ since
it falls in the range where it can be completely determined by the Riemann-Roch
Theorem. To find the exact improvement, the parameters of the best AQECCs derivable 
from one-point codes, \textit{i.e.} codes with divisors $G_1 = iP$ and $G_2 = jP$, were 
stored separately and then compared with the parameters of the AQECCs constructed from 
nested two-point codes. 

Based on the computational data, we present in Table~\ref{Tab:better64} all
two-point codes which strictly improve on one-point codes. 
The resulting quantum codes are of parameters $[511,k,d_{z}/d_{x}]_{64}$.

\begin{table}
  \centering
  \begin{tabular}{@{}llccc@{}}
    \toprule
Best 2-point& Closest 1-point & I& $G_1$ & $G_2$ \\
$(k,d_z,d_x)$& $(k,d_z,d_x)$& &  &  \\
    \midrule
(1, 470, 11)&    (1, 470, 10)    &     1    &$ 35P+5Q $& $35P+ 6Q$\\
(1, 471, 10)&    (1, 470, 10)    &     1    &$ 34P+ 5Q$& $34P+ 6Q$\\
(2, 469, 11)&    (2, 469, 10)    &     1    &$ 35P+ 5Q$& $35P+ 7Q$\\
(2, 470, 10)&    (1, 470, 10)    &     1    &$ 34P+ 5Q$& $34P+ 7Q$\\
(2, 486, 5) &     (2, 486, 4)    &     1    &$ 17P+ 6Q$& $17P+ 8Q$\\
(2, 487, 4) &     (2, 486, 4)    &     1    &$ 16P+ 6Q$& $17P+ 7Q$\\
(3, 460, 14)&    (3, 460, 12)    &     2    &$ 44P+ 4Q$& $44P+ 7Q$\\
(3, 461, 13)&    (3, 460, 12)    &     2    &$ 43P+ 4Q$& $43P+ 7Q$\\
(3, 463, 12)&    (3, 460, 12)    &     3    &$ 41P+ 4Q$& $43P+ 5Q$\\
(3, 477, 7) &     (3, 477, 5)    &     2    &$ 26P+ 5Q$& $26P+ 8Q$\\
(3, 479, 6) &     (3, 477, 5)    &     3    &$ 24P+ 5Q$& $26P+ 6Q$\\
(3, 486, 4) &     (2, 486, 4)    &     1    &$ 16P+ 6Q$& $17P+ 8Q$\\
(4, 462, 12)&    (3, 460, 12)    &     3    &$ 41P+ 4Q$& $43P+ 6Q$\\
(4, 468, 9) &     (4, 468, 6)    &     3    &$ 35P+ 4Q$& $35P+ 8Q$\\
(4, 471, 8) &     (4, 468, 6)    &     5    &$ 32P+ 4Q$& $35P+ 5Q$\\
(4, 478, 6) &     (3, 477, 5)    &     3    &$ 24P+ 5Q$& $26P+ 7Q$\\
(5, 461, 12)&    (3, 460, 12)    &     3    &$ 41P+ 4Q$& $43P+ 7Q$\\
(5, 463, 10)&     (5, 459, 7)    &     7    &$ 40P+ 3Q$& $44P+ 4Q$\\
(5, 470, 8) &     (4, 468, 6)    &     5    &$ 32P+ 4Q$& $35P+ 6Q$\\
(5, 477, 6) &     (5, 476, 5)    &     2    &$ 24P+ 5Q$& $26P+ 8Q$\\
(6, 462, 10)&     (5, 459, 7)    &     7    &$ 40P+ 3Q$& $44P+ 5Q$\\
(6, 469, 8) &     (6, 467, 6)    &     4    &$ 32P+ 4Q$& $35P+ 7Q$\\
(7, 461, 10)&     (7, 458, 7)    &     6    &$ 40P+ 3Q$& $44P+ 6Q$\\
(7, 468, 8) &     (6, 467, 6)    &     4    &$ 32P+ 4Q$& $35P+ 8Q$\\
(8, 460, 10)&     (7, 458, 7)    &     6    &$ 40P+ 3Q$& $44P+ 7Q$\\
(9, 459, 10)&     (7, 458, 7)    &     6    &$ 40P+ 3Q$& $44P+ 8Q$\\
\bottomrule
  \end{tabular}
\caption{Better AQECCs from two-point codes on 
Hermitian curves over $\ff_{64}$. Improvement is measured by adding 
the gain to the closest one-point code in all three parameters $k,d_{x}\text{, and }d_{z}$.}
\label{Tab:better64}
\end{table}

\begin{IEEEbiographynophoto}
  {M.~Frederic} Ezerman is finishing his PhD studies under research scholarship at the Division of Mathematical Sciences at Nanyang Technological University, Singapore. He works mostly on the interplays between classical and quantum error-correcting codes under the guidance of San Ling.
\end{IEEEbiographynophoto}
\begin{IEEEbiographynophoto}
  {Radoslav} Kirov received his PhD in Mathematics from the University of Illinois at Urbana-Champaign, under the supervision of Iwan Duursma. Currently, he is a research fellow at Nanyang Technological University, Singapore.
\end{IEEEbiographynophoto}

\newpage

\begin{thebibliography}{10}
\bibitem{AK01}
A.~Ashikhmin and E.~Knill,
``{Nonbinary quantum stabilizer codes},'' 
\emph{IEEE Trans. Inf. Theory}, vol.~47, no.~7, pp. 3065--3072, 2001.

\bibitem{Bee07}
P.~Beelen, ``The order bound for general algebraic geometric codes,'' {\em
  Finite Fields Appl.}, vol.~13, no.~3, pp.~665--680, 2007.

\bibitem{DuuKirPar10}
I.~Duursma, R.~Kirov, and S.~Park, ``Distance bounds for algebraic geometric
  codes,'', to appear in Journal of Pure and Applied Algebra, 2010.

\bibitem{DuuKir10}
I.~Duursma and R.~Kirov, ``Improved Two-Point Codes on Hermitian Curves'', 
to appear in IEEE: Transactions of Information Theory, 2011.

\bibitem{HomKim05}
M.~Homma and S.~J. Kim, ``Toward the determination of the minimum distance of
  two-point codes on a {H}ermitian curve,'' {\em Des. Codes Cryptogr.},
  vol.~37, no.~1, pp.~111--132, 2005.

\bibitem{HomKim06}
M.~Homma and S.~J. Kim, ``The complete determination of the minimum distance of
  two-point codes on a {H}ermitian curve,'' {\em Des. Codes Cryptogr.},
  vol.~40, no.~1, pp.~5--24, 2006.

\bibitem{KKKS06}
A.~Ketkar, A.~Klappenecker, S.~Kumar, P.~K.~Sarvepalli,
``{Nonbinary stabilizer codes over finite fields},''
\emph{IEEE Trans. Inf. Theory}, vol.~52, pp. 4892--4914, 2006.

\bibitem{Mat01}
G.~L. Matthews, ``Weierstrass pairs and minimum distance of {G}oppa codes,''
  {\em Des. Codes Cryptogr.}, vol.~22, no.~2, pp.~107--121, 2001.

\bibitem{Park10}
S.~Park, ``Minimum distance of Hermitian two-point codes,'' {\em Designs, Codes
  and Cryptography}, vol.~57, no.~2, pp.~195--213, 2010.

\bibitem{SarKla06}
P.~K.~Sarvepalli and A.~Klappenecker 
``{Nonbinary quantum codes from Hermitian curves},''
\emph{Applied algebra, algebraic algorithms and error-correcting codes}, 
 136--143, Lecture Notes in Comput. Sci., 3857, Springer, Berlin, 2006.
\bibitem{SKR09}
P.~K.~Sarvepalli, A.~Klappenecker, and M.~R{\"o}tteler,
``{Asymmetric quantum codes: constructions, bounds and performance},''
\emph{Proc. of the Royal Soc. A (2009)}, vol.~465, pp. 1645--1672, 4 March 2009.

\bibitem{Sti88}
H.~Stichtenoth, ``A note on {H}ermitian codes over {${\rm GF}(q\sp 2)$},'' {\em
  IEEE Trans. Inform. Theory}, vol.~34, no.~5, part 2, pp.~1345--1348, 1988.
\newblock Coding techniques and coding theory.

\bibitem{Sti09}
H.~Stichtenoth, {\em Algebraic function fields and codes}, vol.~254 of {\em
  Graduate Texts in Mathematics}.
\newblock Berlin: Springer-Verlag, second~ed., 2009.

\bibitem{Tie87}
H.~J. Tiersma, ``Remarks on codes from {H}ermitian curves,'' {\em IEEE Trans.
  Inform. Theory}, vol.~33, no.~4, pp.~605--609, 1987.

\bibitem{TsfVla07}
M.~Tsfasman, S.~Vl{\u{a}}du{\c{t}}, and D.~Nogin, {\em Algebraic geometric
  codes: basic notions}, vol.~139 of {\em Mathematical Surveys and Monographs}.
\newblock Providence, RI: American Mathematical Society, 2007.

\bibitem{WFLX09}
L.~Wang, K.~Feng, S.~Ling, and C.~Xing,
``Asymmetric quantum codes: characterization and constructions,''
{\em IEEE Trans. Inform. Theory}, vol.~56, pp. 2938--2945, 2010.

\end{thebibliography}
\end{document}